\documentclass{article} 
\usepackage{nips13submit_e,times}
\usepackage{hyperref}
\usepackage{url}

\usepackage{amsmath, amssymb, amsbsy, amsfonts, bm,subfig,multirow}
\usepackage{xspace}
\usepackage{graphicx}

\makeatletter
\renewcommand*\env@matrix[1][*\c@MaxMatrixCols c]{%
  \hskip -\arraycolsep
  \let\@ifnextchar\new@ifnextchar
  \array{#1}}
\makeatother

\newcommand{\wrt}{\emph{w.r.t.}\xspace}
\newcommand{\eg}{\emph{e.g.}\xspace}
\newcommand{\ie}{\emph{i.e.}\xspace}
\newcommand{\figref}[1]{figure(\ref{#1})\xspace}
\newcommand{\tabref}[1]{table(\ref{#1})\xspace}
\newcommand{\secref}[1]{section(\ref{#1})\xspace}
\newcommand{\secrefs}[2]{sections(\ref{#1} \& \ref{#2})\xspace}
\newcommand{\suppsecref}[1]{section(\ref{#1}) of the supplementary material\xspace}

\title{A Bayesian Residual-Based Test for Cointegration}

\author{
Thomas Furmston \\
Department of Computer Science\\
University College London \\
London, WC1E 6BT \\
\texttt{T.Furmston@cs.ucl.ac.uk} \\
\And
Stephen Hailes \\
Department of Computer Science\\
University College London \\
London, WC1E 6BT \\
\texttt{S.Hailes@cs.ucl.ac.uk} \\
\AND
A. Jennifer Morton \\
Department of Physiology, Development and Neuroscience \\
University of Cambridge \\
Cambridge, CB2 3DY \\
\texttt{ajm41@cam.ac.uk} \\
}

\nipsfinalcopy 

\begin{document}

\maketitle

\begin{abstract}
Cointegration is an important concept in the analysis of non-stationary time-series, giving conditions under which a collection of non-stationary processes has an underlying stationary (cointegration) relationship. In this paper we present the first fully Bayesian residual-based test for cointegration, where we consider the whole space of possible cointegration relationships when testing for the presence of cointegration. We first demonstrate that such a test can be performed exactly in the case where the residual process follows a first-order autoregressive process.  We then extend this test to include more complex residual processes, where we first consider a suitable cointegration test-statistic and then leverage Bayesian sampling techniques to perform the necessary inference. We empirically demonstrate that our Bayesian approach attains a superior classification accuracy than existing approaches, all of which use a point estimate of the cointegration relationship in their test. Finally, we demonstrate our approach on some real world financial time-series data.
\end{abstract}

\section{Introduction}\label{section_autoregrssions_unitRoots_cointegration}

In this section we introduce the theoretical background necessary to understand the concept of cointegration, while also providing a summary of existing techniques in the cointegration literature. Firstly, a univariate $k^{\textnormal{th}}$-order autoregressive process, $X_t$, is defined as  
\begin{equation}
X_t = \phi_1 X_{t-1} + \phi_2 X_{t-2} + \cdots + \phi_k X_{t-k} + \epsilon_t, \label{pthOrderUnivariateAutoregression}
\end{equation}
where $\phi_{1:k} = \begin{bmatrix} \phi_1 & \phi_2 & \cdots & \phi_k \end{bmatrix}^\top \in \mathbb{R}^k$ and the process, $\epsilon_t$, is assumed to be white noise, \ie independent zero-mean Gaussian random variables with constant variance, $\sigma^2 \in \mathbb{R}^+$. It is also possible to include deterministic terms, such as an intercept or a trend term, in (\ref{pthOrderUnivariateAutoregression}). 


Introducing the lag operator, $L(X_t) = X_{t-1}$, an autoregressive process can be written in the equivalent form $\Psi(L)X_t = \epsilon_t$, where $\Psi(L)$ is a polynomial in the lag operator and this polynomial is given by
\begin{equation}
\Psi(z) = 1 - \phi_1 z - \phi_2 z^2 - \cdots - \phi_k z^k. 
\end{equation}
It is well known that an autoregressive process is covariance-stationary (or trend-stationary if a trend term is present in (\ref{pthOrderUnivariateAutoregression})) if the zeros of $\Psi$ lie outside the unit circle, see \eg \cite{hamilton-time-series}. When the context is clear we shall simply refer to a covariance-stationary (or trend-stationary) process as stationary. The class of stationary autoregressive models is rich enough to model many time-series of interest. They are, however, inappropriate for modelling non-stationary time-series and in such cases unit root processes are one of the models of choice. An autoregression is said to follow a unit root process if one of the roots of $\Psi$ is unity and all of the remaining roots lie outside the unit circle. An alternative expression for a unit root process is to say that it is \textit{integrated} of order one, typically denoted by $X_t \sim I(1)$. Extending this notation in the natural manner we use $X_t \sim I(0)$ to denote stationary processes.


A set of $n$ unit root processes, denoted in vector form by $\bm{Z}_t$, is said to be cointegrated if there exists, $\bm{\beta} \in \mathbb{R}^n$, where all elements of $\bm{\beta}$ are non-zero, such that $\bm{\beta}^\top \bm{Z}_t \sim I(0)$. If such a vector exists then it is known as the cointegration vector, or the cointegration relationship. The condition that all the elements of $\bm{\beta}$ are non-zero ensures that non-stationary series are not trivially cointegrated. Cointegration describes the conditions under which a collection of non-stationary processes has an underlying stationary relationship. Cointegration is important because while the individual processes are unpredictable, \ie have no well-defined mean or variance, there is an underlying process that is predictable and can be estimated in a statistically meaningful manner. Cointegration is an important tool in any application that requires the determination of (or modelling of) a relationship between non-stationary time-series. For instance, cointegration plays a prominent role in financial fields, such as econometrics, where groups of stocks, bonds or economic indicators often exhibit an underlying equilibrium relationship. An example of such a relationship is given by the economic indicators of consumption and income. Both indicators are typically modelled as non-stationary processes, but consumption is (on average) a given proportion of income, so that the difference between the logarithm of these two indicators should form a stationary process. Given that this is actually the case then regressing one indicator upon the other will result in a statistically meaningful results. 

There are two fundamentally different approaches to cointegration analysis, namely \textit{residual-based} methods and \textit{error-correction} methods. In \textit{residual-based} methods one component of $\bm{Z}_t$ is regressed upon the remaining components of $\bm{Z}_t$. We use the notation $\bm{X}_t$ and $Y_t$ to respectively denote the regressors and regressand of this regression, where we shall assume \emph{w.l.o.g.} that $Y_t$ forms the first component of $\bm{Z}_t$. Denoting the regression coefficients by $\bm{\beta}_2 \in \mathbb{R}^{n-1}$ then we have that $\bm{\beta}^\top = \begin{bmatrix} 1 & - \bm{\beta}_2^\top \end{bmatrix}$ and  
\begin{equation}
\bm{\beta}^\top \bm{Z}_t = Y_t - \bm{\beta}_2^\top \bm{X}_t = R_t, \label{cointegrationRelationshipEquation}
\end{equation}
where we have used the notation, $R_t$, to denote the residual process. It is common in practice to also include an intercept term in the regression in (\ref{cointegrationRelationshipEquation}), which is equivalent to including a constant term in the residual process. In this case we shall write $R_t = Y_t - \bm{\beta}_2^\top \bm{X}_t - \alpha$, where $\alpha \in \mathbb{R}$ is the intercept term. Cointegration analysis now amounts to determining the stationarity properties of the residual process that results from this linear regression. We shall sometimes use the notation $R_t(\bm{\beta}_2)$ to denote the residual process induced from the regression coefficients, $\bm{\beta}_2\in \mathbb{R}^{n-1}$. In \textit{error-correction} methods the entire multivariate cointegrated system is modelled directly through an error-correction model \cite{engle-1987}. The cointegration space (\ie the space of cointegrating vectors) is then estimated through the reduced-rank long term matrix of the error-correction model. There are advantages and disadvantages to both approaches, but in this paper our interest is on performing regression analysis between non-stationary time-series and for this reason we focus on residual-based methods. 


The standard frequentist approach to residual-based cointegration testing is the Engle-Granger method \cite{engle-1987}. The Engle-Granger method is a two stage test, where the first stage consists of obtaining a point estimate of the regression coefficients, $\hat{\bm{\beta}}_2 \in \mathbb{R}^{n-1}$, while the second stage consists of testing the stationarity of $R_t(\hat{\bm{\beta}}_2)$. This point estimate can be obtained through any number of techniques, but in practice it is typically obtained through linear regression. The stationarity of $R_t(\hat{\bm{\beta}}_2)$ is tested through a standard unit root hypothesis test, such as the augmented Dickey-Fuller test \cite{dickey_fuller_1979}, where the null hypothesis is that $R_t(\hat{\bm{\beta}}_2)$ has a unit root and the alternative hypothesis is that it is stationary. If the null hypothesis is rejected then it is concluded that the collection of time-series are cointegrated. Bayesian approaches to cointegration analysis are almost exclusively focused on error-correction techniques, see \eg \cite{koop2004} and references therein, with the exception of the recent paper \cite{bracegirdle-2012-icml}. The method of \cite{bracegirdle-2012-icml} is like the Engle-Granger method in the sense that a point estimate of the regression coefficients is first constructed, again denoted by $\hat{\bm{\beta}}_2 \in \mathbb{R}^{n-1}$, and then the stationarity of $R_t(\hat{\bm{\beta}}_2)$ is tested. The method of \cite{bracegirdle-2012-icml} differs from the Engle-Granger method in two ways. Firstly, an unbiased point estimate is obtained by optimising the likelihood over the space of stationary models, where the EM-algorithm \cite{dempster-1977} is used to perform this optimisation. Secondly, a Bayesian unit root test \cite{schotman-1991} is used to test the stationarity of $R_t(\hat{\bm{\beta}}_2)$, where the Bayes' factor is used to construct this test. Due to the fact that a point estimate of the regression coefficients is used in \cite{bracegirdle-2012-icml} we refer to this method as a partially Bayesian residual-based test for cointegration. 

In this paper we propose a fully Bayesian residual-based test for cointegration. In particular, we place a prior over the regression coefficients and we then marginalise out this variable when constructing a test for detecting cointegration. We thus consider the entire space of possible regression coefficients when testing for cointegration, as opposed to a single point estimate as in the Engle-Granger method or the partially Bayesian cointegration test.



\section{First-Order Autoregressive Residual Processes}\label{section_AR(1)_cointegration_test}

In this section we shall construct two Bayesian tests for cointegration. We currently assume that the residual process follows a first-order autoregressive process, where we shall relax this constraint in \secrefs{section_AR(p)_cointegration_test_p_known}{section_AR(p)_cointegration_test_p_unknown}. As the residual process is assumed to follow a first-order autoregressive process we have that
\begin{equation}
R_t = \phi R_{t-1} + \epsilon_t, \label{firstOrderResidualAutoregressiveProcess}
\end{equation}
where $\phi \in \mathbb{R}$ and $\epsilon_t$ is a white noise process. Combining (\ref{cointegrationRelationshipEquation}) and (\ref{firstOrderResidualAutoregressiveProcess}) means that for all $t \ge 2$ we have 
\begin{equation}
Y_t = \bm{\beta}_2^\top \bm{X}_t + \phi \big( Y_{t-1} - \bm{\beta}_2^\top \bm{X}_{t-1} \big)  + \epsilon_t, \label{AR1LikelihoodTerm}
\end{equation}
where the distributional form of the initial observation depends on how we model the initial observation of the residual process. The modelling of the initial observation is a non-trivial problem and is complicated by factors such as the non-existence of a well-defined distribution for the initial observation of a unit root process, as well as the constraints imposed on this distribution in the case of a stationary process. We shall return to this point when we consider how to perform posterior inference of the autoregression parameter in this model. As we are taking a Bayesian perspective we also place priors on the regression coefficients, the autoregression parameter and the variance of the white noise process, where we shall consider the prior $p(\bm{\beta}_2, \phi, \sigma^2) \propto \sigma^{-2}$.

Given a realisation of the system, which we denote by $\bm{z}_{1:T}$, we wish to determine whether or not the system is cointegrated. In the residual-based framework this is equivalent to determining whether or not the residual process in (\ref{cointegrationRelationshipEquation}) is stationary for some $\bm{\beta}_2 \in \mathbb{R}^{n-1}$. As each of the components of $\bm{Z}_t$ is $I(1)$ it follows that either $R_t \sim I(0)$ or $R_t \sim I(1)$, so it is sufficient to determine whether or not there is a unit root present in the residual process. For the autoregression (\ref{firstOrderResidualAutoregressiveProcess}) the form of $\Psi$ is given by $\Psi(z) = 1 - \phi z$, so there is a unit root in (\ref{firstOrderResidualAutoregressiveProcess}) if $\phi = 1$, while it is stationary if $|\phi| < 1$. These are the criteria that we shall use in our Bayesian tests for cointegration, where we shall consider two such tests. The first test we consider uses Bayes' factors, where the Bayes' factor is given by the ratio of the conditional marginal likelihood under the unit root model with the conditional marginal likelihood under the stationary model, \ie 
\begin{equation}
K = \frac{ p(Y_{2:T} = y_{2:T}| \bm{X}_{1:T}=\bm{x}_{1:T}, Y_1 = y_1, \phi=1)}{\frac{1}{2} \int_{-1}^1 \ d \phi \ p(Y_{2:T} = y_{2:T}| \bm{X}_{1:T}=\bm{x}_{1:T}, Y_1 = y_1, \phi)}. \label{bayesFactorVersion1}
\end{equation}
The likelihood terms in (\ref{bayesFactorVersion1}) are obtained by integrating out the regression coefficients and the variance of the white noise process with respect to their respective priors. When $K \ge \alpha$, for some threshold $\alpha \in \mathbb{R}^+$, then we conclude that the system $\bm{Z}_t$ is not cointegrated, otherwise we conclude that it is cointegrated. We consider the conditional likelihood in (\ref{bayesFactorVersion1}) as this obviates the need to construct a consistent prior for the initial observation of the residual process, which, as previously mentioned, is not possible. The conditional likelihood is an approximation to the full likelihood, but asymptotically the two will give equivalent results. The second test we consider, which is analagous to methods used in the Bayesian unit root literature \cite{phillips-1991,Lubrano-1995}, is to construct the marginal posterior of $\phi$ and then to test whether the posterior probability mass of a stationary model exceeds a certain threshold. In our test this amounts to testing whether $p( |\phi| < 1| \bm{Z}_{1:T} = \bm{z}_{1:T}) \ge 1 - \alpha$ (or equivalently $p( \phi \ge 1| \bm{Z}_{1:T} = \bm{z}_{1:T}) \le \alpha$), for some threshold, $\alpha \in [0,1]$, and rejecting the hypothesis of no cointegration when this condition is satisfied. We shall refer to this test as a credible interval test.

To perform either of these Bayesian cointegration tests it necessary to calculate either the conditional marginal likelihood, $p(Y_{2:T} = y_{2:T}| \bm{X}_{1:T}=\bm{x}_{1:T}, Y_{1} = y_{1}, \phi)$, or the marginal posterior, $p( \phi| \bm{Z}_{1:T} = \bm{z}_{1:T})$. As we are considering a flat prior on $\phi$ these two terms are proportional to each other \wrt $\phi$, and so we detail the calculation of the marginal likelihood. This calculation is complicated by the fact that when there is an intercept present in the regression the model is locally non-identifiable \wrt the intercept. In particular, we have in this case 
\begin{equation}
Y_t = \bm{\beta}_2^\top \big( \bm{X}_t - \phi \bm{X}_{t-1} \big) + \phi Y_{t-1} + (1-\phi) \alpha + \epsilon_t,
\end{equation}
which doesn't depend on the intercept when $\phi = 1$. This local non-identifiability will lead to a divergent Bayes' factor, or an ill-defined posterior, when a non-informative prior is placed on the intercept and the conditional marginal likelihood is considered. One possible solution to this problem would be to use an informative prior, but this requires some prior information about the intercept and this may not be available in practice. Instead we shall tackle this problem by placing a suitable prior over the initial observation of the residual process and considering the full marginal likelihood, \ie $p(Y_{1:T} = y_{1:T}| \bm{X}_{1:T}=\bm{x}_{1:T}, \phi)$. 

In the case of the Bayes' factor we consider the prior $p(R_1|\phi, \sigma^2) = \mathcal{N}(0, \sigma^2/(1-\phi^2))$ for $\phi \in (-1,1)$, where this prior enforces the stationarity constraints on the residual process \cite{hamilton-time-series}. Under this prior the full marginal likelihood takes the form  
\begin{equation}
p(Y_{1:T} = y_{1:T}|\bm{X}_{1:T} = \bm{x}_{1:T}, \phi) \propto \bigg( \frac{ 1 - \phi^2  }{ g(\phi)^{T-n} \big| L_{XX}(\phi) \big| } \bigg)^{\frac{1}{2}}. \label{likelihood_ar1_prior3_full_likelihood}
\end{equation}
We use the proportional notation to denote that there are multiplicative factors that are independent of $\phi$. The notation in (\ref{likelihood_ar1_prior3_full_likelihood}) is  
\begin{align*}
& \quad \quad \quad \quad  g(\phi) = L_{YY}(\phi) - L^\top_{XY} (\phi) \big( L_{XX}(\phi) \big)^{-1} L_{XY}(\phi), \\
L_{XX}(\phi) &= (1 - \phi^2) \begin{bmatrix} 1 & \bm{x}_1^\top \\ \bm{x}_1 & \bm{x}_1 \bm{x}_1^\top \end{bmatrix}    + \sum_{t=2}^T \bigg( \begin{bmatrix} 1 \\ \bm{x}_t \end{bmatrix} - \phi \begin{bmatrix} 1 \\ \bm{x}_{t-1} \end{bmatrix} \bigg) \bigg( \begin{bmatrix} 1 \\ \bm{x}_t \end{bmatrix} - \phi \begin{bmatrix} 1 \\ \bm{x}_{t-1} \end{bmatrix} \bigg)^\top,
\end{align*}
where the terms $L_{XY}(\phi)$ and $L_{YY}(\phi)$ are defined in a similar manner. See \suppsecref{supp_section_AR(1)_cointegration_test} for more details. This prior is not well defined for the unit root process and so we calculate the likelihood in this case by taking the limit of the likelihood for stationary models, \ie $p(Y_{1:T} = y_{1:T}|\bm{X}_{1:T} = \bm{x}_{1:T}, \phi=1) = \lim_{\phi \to 1} p(Y_{1:T} = y_{1:T}|\bm{X}_{1:T} = \bm{x}_{1:T}, \phi)$. Details on the conditions for the existence of this limit are given in  \suppsecref{supp_section_AR(1)_cointegration_test}. In the case of the test based on credible intervals, where we necessarily consider a prior over the whole space of autoregressive models, we consider the prior $p(R_1|\phi, \sigma^2) = \mathcal{N}(0, \sigma^2)$ and the calculation of the full likelihood follows in an analagous manner. Additionally, the calculation of the conditional likelihood in the absence of an intercept can also be performed using similar calculations.

\section{Higher-Order Autoregressive Residual Processes}\label{section_AR(p)_cointegration_test_p_known}
In \secref{section_AR(1)_cointegration_test} we considered the case where the residual process was modelled through a first order autoregressive process. However, in many cases of interest it may happen that the correlation structure of the residual process is too complicated to be properly captured through this model and a higher order autoregressive process would be more appropriate. In this section we construct a residual-based cointegration test for the case where the residual process is modelled through a $k^{\textnormal{th}}$-order autoregressive process, for general $k \in \mathbb{N}$. We assume that the order of the residual process is known, where we shall relax this constraint in \secref{section_AR(p)_cointegration_test_p_unknown}. In the first part of this section we construct a test-statistic that can be used to test whether or not there is a unit root present in the residual process. The second part of this section then details how to perform marginal posterior inference of this test-statistic. Due to the introduction of nuisance parameters exact inference is no longer possible and so we construct a Gibbs sampler to perform the marginal posterior inference.

As the residual process is assumed to follow a $k^{\textnormal{th}}$-order autoregressive process then we have that 
\begin{equation}
R_t = \phi_1 R_{t-1} + \phi_2 R_{t-2} + \cdots + \phi_k R_{t-k} + \epsilon_t. \label{pthOrderResidualAutoregressiveProcess}
\end{equation}
In order to obtain a test for cointegration in this case it is necessary to construct a test-statistic that can be used to determine whether the residual process is stationary or contains a unit root. Recall that the residual process will be stationary when all the roots of $\Psi$ lie outside the unit circle, while it will be a unit root process when one of the roots is unity and the remaining roots lie outside the unit circle. Now to obtain a test-statistic we rewrite (\ref{pthOrderResidualAutoregressiveProcess}) into the form
\begin{equation}
R_t = \rho R_{t-1} + \xi_1 \Delta R_{t-1} + \xi_2 \Delta R_{t-2} + \cdots + \xi_{k-1} \Delta R_{t-(k-1)} + \epsilon_t, \label{pthOrderResidualAutoregressiveProcessErrorCorrectionForm} 
\end{equation}
where 
\begin{equation*}
\rho = \sum_{i=1}^k \phi_i, \quad \quad \quad \quad \xi_i = - \sum_{j=i+1}^k \phi_j.
\end{equation*}
The notation $\Delta$ is used to denote first differences, \ie $\Delta R_{t} = R_t - R_{t-1}$. The purpose of writing the residual process in this form is that the parameter for the first lagged level, namely $\rho$, can be written in terms of the reciprocals of the roots of $\Psi$. In particular we have that
\begin{equation}
\rho = (-1)^{p+1} \prod_{i=1}^k (\lambda_i - 1) + 1, \label{rhoInTermsOfZeros}
\end{equation}
where $\{\lambda_i\}_{i=1}^k$ are the roots of the polynomial $\Pi(z) = z^k - \phi_1 z^{k-1} - \phi_2 z^{k-2} - \cdots - \phi_k.$ Note that the roots of $\Psi$ are the reciprocals of the roots of $\Pi$. Details of the derivations of (\ref{pthOrderResidualAutoregressiveProcessErrorCorrectionForm}) and (\ref{rhoInTermsOfZeros}) are given in \suppsecref{supp_section_AR(p)_cointegration_test_p_known}. It can be seen from (\ref{rhoInTermsOfZeros}) that when there is a unit root present in the residual process, \ie $\lambda_i = 1$ for some $i \in \mathbb{N}_k$, then $\rho=1$. Conversely, when the residual process is stationary, so that $|\lambda_i| < 1$ for each $i \in \mathbb{N}_k$ and complex roots of $\Pi$ occur in conjugate pairs, we have that $\rho < 1$. Hence to test whether $R_t \sim I(0)$ or $R_t \sim I(1)$ then it is sufficient to test whether $\rho < 1$ or $\rho = 1$, and this is the test-statistic that we shall use in our tests for cointegration.

We shall see shortly that when the residual process is modelled through a $k^{\textnormal{th}}$-order autoregressive processes, for general $k \in \mathbb{N}$, that posterior inference in our cointegration model is no longer tractable. Instead we shall employ sampling methods to perform the inference, where, in particular, we shall construct a Gibbs sampler. For this reason we shall no longer consider using Bayes' factors to test for cointegration, and shall instead consider only the second type of test. This means that in order to test for cointegration it is necessary to calculate the marginal posterior
\begin{equation*}
p(\rho|\bm{Z}_{1:T} = \bm{z}_{1:T}) \propto \int \ d\bm{\xi} \ d \sigma^2 \ d \bm{\beta}_2 \ p(Y_{k+1:T} = y_{k+1:T} | \bm{X}_{1:T} = \bm{x}_{1:T}, \rho, \bm{\xi}, \sigma^2, \bm{\beta}_2) p(\bm{\xi}, \sigma^2, \bm{\beta}_2),
\end{equation*}
where we are considering the case where no intercept is included in the regression, and so are considering the conditional likelihood. The methods of this section can be easily extended to the case where an intercept is present by using analagous methods to those presented in \secref{section_AR(1)_cointegration_test}. Given the form of the residual process (\ref{pthOrderResidualAutoregressiveProcessErrorCorrectionForm}) and the cointegration relationship (\ref{cointegrationRelationshipEquation}) we have
\begin{equation}
Y_t = \bm{\beta}^\top_2 \bm{X}_t + \rho \big( Y_{t-1} - \bm{\beta}^\top_2 \bm{X}_{t-1} \big) + \sum_{i=1}^{k-1} \xi_i \big( \Delta Y_{t-i} - \bm{\beta}^\top_2 \Delta \bm{X}_{t-i} \big) + \epsilon_t, \quad \quad t \ge k+1, \label{YLikelihoodTermHigherOrder}
\end{equation}
where we have used the fact that $\Delta R_t = \Delta Y_t - \bm{\beta}_2^\top \Delta \bm{X}_t$. We place a flat prior over the parameters of the autoregression. As in \secref{section_AR(1)_cointegration_test} we consider the prior $p(\sigma^2) \propto \sigma^{-2}$ for the variance of the white noise process.

The introduction of the nuisance parameters $\bm{\xi}$ that occurs when considering a residual process of the form (\ref{pthOrderResidualAutoregressiveProcess}) precludes exact posterior inference. This can be seen by the fact that the exponent of the likelihood term (\ref{YLikelihoodTermHigherOrder}) is not jointly quadratic in $\bm{\xi}$ and $\bm{\beta}_2$, which means that it is not possible to marginalise out both of these variables in closed form. While not jointly quadratic in $\bm{\xi}$ and $\bm{\beta}_2$ the exponent of (\ref{YLikelihoodTermHigherOrder}) is quadratic in each of these variables individually, considering the other variable as fixed. This suggests that Gibbs sampling is appropriate and this is the approach that we take here, iteratively sampling from the following conditional distributions
\begin{align*}
p(\rho, \bm{\xi}|\bm{Z}_{1:T}=\bm{z}_{1:T}, \sigma^2, \bm{\beta}_2), \quad  \quad  p(\bm{\beta}_2|\bm{Z}_{1:T}=\bm{z}_{1:T}, \rho, \bm{\xi}, \sigma^2), \quad \quad  p(\sigma^2|\bm{Z}_{1:T}=\bm{z}_{1:T}, \rho, \bm{\xi}, \bm{\beta}_2).
\end{align*}
The first two distributions are Gaussian, while the third is an inverse Gamma distribution. Due to reasons of space we give the form of these conditional distributions in \suppsecref{section_AR(p)_cointegration_test_p_known}. As these conditional distributions are of standard form they can be sampled from efficiently and a Gibbs sampler can be implemented in a standard manner.


\section{Autoregressive Residual Processes with Unknown Order}\label{section_AR(p)_cointegration_test_p_unknown}
Until now we have considered the case where the order of the residual process is assumed to be known. However, this is typically not the case in practice and so some form of model selection is required. One of the standard Bayesian approaches to model selection is to index the various models, considering this index variable as a random variable, and then to obtain the posterior over this index variable. In the case where there are a different number of variables in the various models posterior inference is typically done through reversible jump Markov chain Monte Carlo (RJ-MCMC) methods \cite{green-1995}. As we are modelling the residual process through an autoregressive process performing model selection requires the construction of a RJ-MCMC sampling algorithm that jumps between autoregressive models of different order. RJ-MCMC sampling methods have previously been considered for autoregressive models, see \emph{e.g.} \cite{Troughton97,Ehlers06,brooks-2003}, and we follow an analagous framework here. We shall consider the case where the order of the residual process takes on the possible values, $k \in \{ 0, 1, ..., k_{\textnormal{max}} \}$, for some $k_{\textnormal{max}} \in \mathbb{N}$, where we consider a uniform prior for this variable. 

In our RJ-MCMC framework there are two different types of moves in the parameter space, \textit{within-model moves} and \textit{between-model moves}. In a within-model move the order of the residual process is held fixed and the parameters of the model are sampled, where this is done through the Gibbs sampler presented in \secref{section_AR(p)_cointegration_test_p_known}. In a between-model move the order of the residual process is sampled, where this requires a reversible jump move as it necessarily involves a change in the dimension of the parameter space. The first step of a between-model move is to propose a move from order $p$ to order $p'$, then given this new proposed model order a new vector of autoregressive parameters, $\phi'_{1:k'} \in \mathbb{R}^{k'}$, is proposed. We hold the regression coefficients and the variance of the white noise process fixed during between-model moves, but this is not necessary in practice. Given the proposed $k'$ and $\phi'_{1:k'}$ the move is accepted with probability $\textnormal{min}\{1, A((k, \phi_{1:k}) \to (k', \phi'_{1:k'})) \}$, where $A((k, \phi_{1:k}) \to (k', \phi'_{1:k'}))$ is the standard acceptance ratio for accepting the proposed move, see \eg \cite{green-1995,Troughton97} for more details.

We propose moves in the order of the residual process by using the discretised Laplacian density, where this is given by $q(k'|k) \propto \exp \big( - \lambda |k' - k| \big)$. The parameter, $\lambda \in \mathbb{R}^+$, is a `heat' parameter that determines the spread of the distribution. Given the proposed new model order, the regression coefficients and the variance of the white noise process the conditional distribution of the autoregressive parameters can be obtained analytically, as shown in \secref{section_AR(p)_cointegration_test_p_known}. We use this conditional distribution as the proposal distribution of the autoregressive parameters in the new model. Under these proposal distributions the acceptance ratio can be calculated using the \textit{Candidate's identity} \cite{besag-1989}. This allows the acceptance ratio to be calculated in such a manner that it is only necessary to sample the autoregression parameters once a move has been accepted. The derivation of the acceptance ratio is straightforward, but algebraically cumbersome, and so we give the details in \suppsecref{supp_section_AR(p)_cointegration_test_p_unknown}. 

Posterior inference can be performed by iteratively sampling in the standard manner, alternating between within-model moves and between-model moves. The marginal posterior of $\rho$ can be obtained by constructing the conditional marginal of $\rho$, conditioned on the order of the residual process, and then marginalizing out the order of the residual process \emph{w.r.t.} its posterior distribution. This marginal posterior can then be used to test for cointegration in the same manner as in \secref{section_AR(p)_cointegration_test_p_known}. Furthermore, given that the observed time-series are deemed to be cointegrated, then the sample regression coefficients can be used to obtain the posterior of the cointegration relationship.

\section{Experiments}\label{section_experiments}
In this section we evaluate the various methods proposed in this paper on both synthetic and real world data. We begin by constructing the \textit{receiver operating characteristic} (ROC) curves in order to determine the classification accuracy of the methods proposed in \secrefs{section_AR(1)_cointegration_test}{section_AR(p)_cointegration_test_p_known}. We then evaluate the accuracy of the RJ-MCMC algorithm of \secref{section_AR(p)_cointegration_test_p_unknown} in determining the model structure of the unobserved residual process. Finally, we demonstrate the method on some real world financial time-series data.

\begin{figure}[t]
\centering
\includegraphics[width=5in, height=2in]{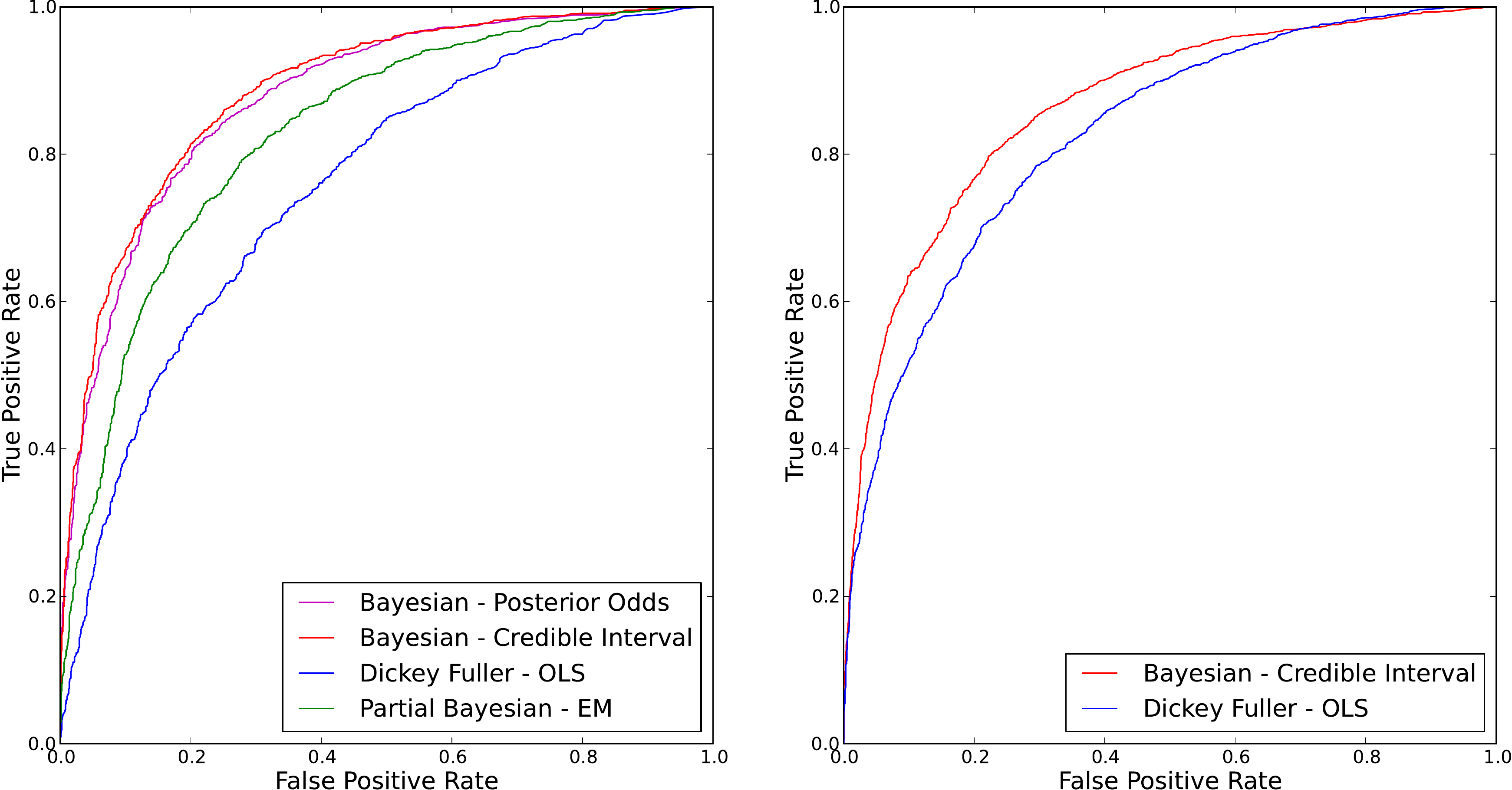}\label{fig:AR1_ROC_Curve}
\caption{Results of the residual-based cointegration classification test, where the plots show the ROC curve for the experiment with (a) a first-order autoregressive residual process and (b) a third-order autoregressive residual process. The plots show the \textit{true positive} rate against the \textit{false positive rate} for each of the algorithms, where optimal classification is achieved with a true positive rate of one and a false positive rate of zero.} \label{fig:AR_curve_results} 
\end{figure}

In the first experiment we considered the classification accuracy of the Bayesian residual-based cointegration tests presented in this paper, that is the accuracy with which these tests determine whether a collection of time-series are cointegrated or not. In this experiment we considered the case where the underlying model of the residual process was known \textit{a priori}, so that we considered the tests of \secrefs{section_AR(1)_cointegration_test}{section_AR(p)_cointegration_test_p_known}. For comparison we also considered the Engle-Granger test and the partially Bayesian cointegration test of \cite{bracegirdle-2012-icml}. The test of \cite{bracegirdle-2012-icml} is only applicable when the residual process follows a first-order autoregression, so we were only able to make a comparison with this test in this one particular case. The same applies to our Bayesian test based on the Bayes' factor presented in \secref{section_AR(1)_cointegration_test}. In order to assess the accuracy of the tests we considered synthetically generated data, where the details of the procedure used to generate the data are given in \suppsecref{supp_material_experiments}. We considered two different experiments, where in the first the residual process follows a first-order autoregression, while in the second it follows a third-order autoregression. In both experiments we considered $2,500$ independently generated tests and the results are presented in \figref{fig:AR_curve_results}. The results are presented in the form of a \textit{receiver operating characteristic} curves, where the false positive rate of each test procedure is plotted against its true positive rate. The optimal classification rule would result in a false positive rate of zero and a true positive rate of one, which corresponds to the top left-hand corner of the plot. In can be seen in \figref{fig:AR_curve_results} that the superior classification rate of the Bayesian tests presented in this paper is marked. It can also be seen that the difference in performance between the two types of Bayesian test presented in \secref{section_AR(1)_cointegration_test} is negligible. 

In the second experiment we investigated the accuracy of the RJ-MCMC algorithm of \secref{section_AR(p)_cointegration_test_p_unknown} in determining the model structure of $R_t$. For comparison we also considered the Engle-Granger method, where the \textit{Bayesian information criterion} (BIC) \cite{Schwarz-1978} was used in the unit root test to determine the model of $R_t$. We also considered other standard frequentist model selection techniques, but by comparison to the BIC they gave inferior results in this experiment. In order to assess the accuracy of the RJ-MCMC algorithm we considered synthetically generated data, where the procedure used to generate the data is detailed in \suppsecref{supp_material_experiments}. In the experiment we considered observation sequences of length $T = 100$, $200$, $500$, $750$ and $1000$. Given $\bm{X}_{1:T}$ and $Y_{1:T}$, where $Y_{1:T}$ was regressed upon $\bm{X}_{1:T}$, we used the RJ-MCMC sampling algorithm presented in \secref{section_AR(p)_cointegration_test_p_unknown} to estimate the posterior of the model order of the residual process. We used the mode of the posterior to classify the model order. We also calculated the variance of the posterior in order to obtain a gauge of the dispersion of the posterior. The results of the experiment are given in \tabref{modelOrderExperimentResults}, where the results were obtained from 250 independently generated tests. It can be seen that regardless of the number of observations the classification accuracy of the RJ-MCMC approach was superior to that of the BIC approach. Additionally, a measure of uncertainty in the prediction is given by the RJ-MCMC approach, where it can be seen from \tabref{modelOrderExperimentResults} that when there are fewer observations the posterior is more dispersed.
\begin{table}[b]
\centering
\begin{tabular}{cc|c|c|c|c|c|}
\cline{3-7}
& & \multicolumn{5}{ c| }{\# Observations} \\ \cline{3-7}
& & 100 & 200 & 500 & 750 & 1000 \\ \cline{1-7}
\multicolumn{1}{ |c| }{\multirow{2}{*}{RJ-MCMC} } &
\multicolumn{1}{ |c| }{Accuracy (mode)} & 83.2\% & 90.4\% & 93.2\% & 93.6\% & 94.8\%    \\ \cline{2-7}
\multicolumn{1}{ |c  }{}                        &
\multicolumn{1}{ |c| }{Variance} & 0.15 & 0.09 & 0.05 & 0.04 & 0.03    \\ \cline{1-7}
\multicolumn{2}{ |c| }{Bayesian Information Criterion} & 68.4\% & 78.4\% & 80.0\% & 82.0\% & 81.2\% \\ \cline{1-7}
\end{tabular}
\caption{The results of the experiment to assess the accuracy of the RJ-MCMC algorithm in determining the model of the residual process. The table gives the classification accuracy of the RJ-MCMC algorithm and the Engle-Granger algorithm, using the BIC, for varying numbers of observations. The mode of the posterior was used as a classification rule when using the RJ-MCMC algorithm. The variance of the posterior is also presented. }\label{modelOrderExperimentResults}
\end{table}

\begin{figure}[t]
\centering
\includegraphics[scale=0.39]{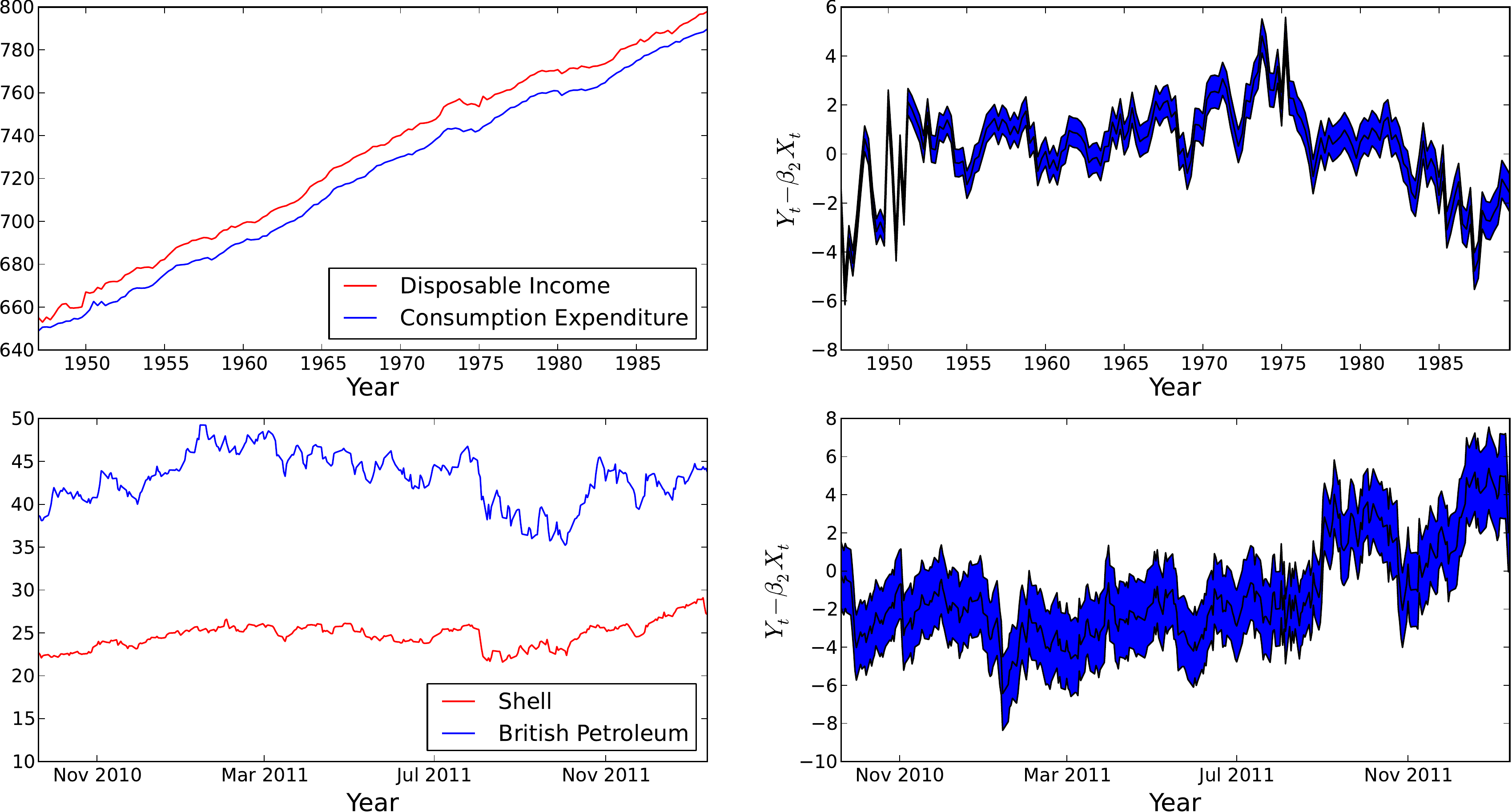}
\caption{Results from the Bayesian residual-based cointegration analysis of the economic indicators for US consumption. The top left plot displays $100$ times the logarithm of the real quarterly aggregate personal disposable income (red) and personal consumption expenditures (blue) for the United States from the first quarter of 1947 through until the third quarter of 1989. The bottom left plot shows the daily stock price of British Petroleum (blue) and Shell (red) from the $20^{\textnormal{th}}$ of September 2010 until the $1^{\textnormal{st}}$ of Janurary 2012. The top right and bottom right plots show the respective posteriors of the residual process, where this is obtained by taking the expectation of $Y_t - \beta_2 X_t$ \emph{w.r.t.} the marginal posterior of the regression coefficient. The plot shows the mean (black) and the three times the standard deviation (blue).}\label{fig-USConsumptionBPShell}
\end{figure}

In the final experiment we applied our Bayesian residual-based cointegration analysis model to two real world financial data sets. The first data set we considered was the economic indicators for personal disposable income and personal consumption expenditure for the United States of America from the first quarter of 1947 through until the third quarter of 1989. According to the economic model of \cite{davidson-1978} expenditure should on average be a certain proportion of income, so that the logarithm of these two indicators should be cointegrated. The second data set we considered was the daily stock prices of British Petroleum (BP) and Shell from the $20^{\textnormal{th}}$ of September 2010 until the $1^{\textnormal{st}}$ of Janurary 2012. As these two stock prices relate to companies in the same industry we expect that the stock prices should be cointegrated, at least over a relatively short period of time where no anomalous events occurred. These two data sets, along with the posterior of the residuals obtained from our Bayesian model, are given in \figref{fig-USConsumptionBPShell}. The results from our Bayesian cointegration model strongly indicate that both of these data sets are cointegrated, where in the first data set all of the posterior probability mass was placed on stationary models, while $98\%$ of the posterior mass was placed on stationary models in the second data set. 


\section{Summary}
In this paper we have presented the first fully Bayesian approach to residual-based cointegration analysis, where we have presented a series of tests that are applicable to residual processes of any given (possibly unknown) order. One of the advantages of this Bayesian approach is the ease with which the model can be extended to incorporate more complex data, such as a non-stationary cointegration relationship or intermittent cointegration, through modern Bayesian sampling techniques.

\bibliography{bib_file}
\bibliographystyle{plain}

\newpage 

\section{Supplementary Material}

\subsection{First-Order Autoregressive Residual Processes}\label{supp_section_AR(1)_cointegration_test}

We now show that under the prior, $p(R_1|\phi, \sigma^2) = \mathcal{N}(0, \sigma^2/(1-\phi^2))$, where $\phi \in (-1,1)$, that the full marginal likelihood, takes the form 
\begin{equation}
p(Y_{1:T} = y_{1:T}|\bm{X}_{1:T} = \bm{x}_{1:T}, \phi) \propto \bigg( \frac{ 1 - \phi^2  }{ g(\phi)^{T-n} \big| L_{XX}(\phi) \big| } \bigg)^{\frac{1}{2}}. \label{supp_likelihood_ar1_prior3_full_likelihood}
\end{equation}
The proportional notation is used to denote that there are multiplicative factors that are independent of $\phi$. The notation in (\ref{supp_likelihood_ar1_prior3_full_likelihood}) is  
\begin{align*}
& \quad \quad \quad \quad  g(\phi) = L_{YY}(\phi) - L^\top_{XY} (\phi) \big( L_{XX}(\phi) \big)^{-1} L_{XY}(\phi), \\
L_{XX}(\phi) &= (1 - \phi^2) \begin{bmatrix} 1 & \bm{x}_1^\top \\ \bm{x}_1 & \bm{x}_1 \bm{x}_1^\top \end{bmatrix}    + \sum_{t=2}^T \bigg( \begin{bmatrix} 1 \\ \bm{x}_t \end{bmatrix} - \phi \begin{bmatrix} 1 \\ \bm{x}_{t-1} \end{bmatrix} \bigg) \bigg( \begin{bmatrix} 1 \\ \bm{x}_t \end{bmatrix} - \phi \begin{bmatrix} 1 \\ \bm{x}_{t-1} \end{bmatrix} \bigg)^\top,
\end{align*}
where the terms $L_{XY}(\phi)$ and $L_{YY}(\phi)$ are likewise given by 
\begin{align*}
L_{XY}(\phi) = (1 - \phi^2) y_1 & \begin{bmatrix} 1 \\ \bm{x}_1 \end{bmatrix} + \sum_{t=2}^T \big( y_t - \phi y_{t-1} \big) \bigg( \begin{bmatrix} 1 \\ \bm{x}_t \end{bmatrix} - \phi \begin{bmatrix} 1 \\ \bm{x}_{t-1} \end{bmatrix} \bigg), \\
L_{YY}(\phi) &= (1 - \phi^2) y_1^2 + \sum_{t=2}^T \big( y_t - \phi y_{t-1} \big)^2.
\end{align*}

It can be seen that for each time-point, $t \ge 2$, the likelihood term (\ref{AR1LikelihoodTerm}) can be written in the form  
\begin{equation}
Y_t - \phi Y_{t-1} = \bm{\beta}_2^\top \big( \bm{X}_t - \phi \bm{X}_{t-1} \big) + \alpha (1-\phi) + \epsilon_t,
\end{equation}
while the prior for the initial observation of the residual process gives  
\begin{equation}
Y_1 = \bm{\beta}_2^\top \bm{X}_1 + \alpha + \epsilon_1,
\end{equation}
where $\epsilon_1 \sim \mathcal{N}(0, \sigma^2/(1 - \phi^2))$. This means that the marginal likelihood takes the form 
\begin{align*}
p(Y_{1:T} = y_{1:T}| \bm{X}_{1:T}=\bm{x}_{1:T}, \phi) &= \int \ d \alpha \ d \bm{\beta}_2 \ d \sigma^2 \frac{1}{\sigma^2} \mathcal{N}( y_1 | \bm{\beta}^\top_2 \bm{x}_1 + \alpha, \sigma^2/(1-\phi^2)) \\
&\times \prod_{t=2}^T \mathcal{N}( y_t - \phi y_{t-1} | \bm{\beta}^\top_2 \big( \bm{x}_t - \phi \bm{x}_{t-1} \big) + (1-\phi) \alpha, \sigma^2).
\end{align*}
By expanding the exponent this can be written in the form 
\begin{align*}
p(Y_{1:T} = y_{1:T}| \bm{X}_{1:T}=\bm{x}_{1:T}&, \phi) = \sqrt{1 - \phi^2} \int \ d \alpha \ d \bm{\beta}_2 \ d \sigma^2 \frac{1}{\sigma^2} \bigg( \frac{1}{\sqrt{2 \pi \sigma^2}} \bigg)^T \\
& \times  \exp \bigg( - \frac{1}{2 \sigma^2} \bigg( L_{YY}(\phi) - 2 \begin{bmatrix} \alpha \\ \bm{\beta}_2\end{bmatrix}^\top L_{XY}(\phi) + \begin{bmatrix} \alpha \\ \bm{\beta}_2\end{bmatrix}^\top L_{XX}(\phi) \begin{bmatrix} \alpha \\ \bm{\beta}_2\end{bmatrix} \bigg) \bigg).
\end{align*}
Using standard Gaussian integral formulae we have that the integral \wrt $\alpha$ and $\bm{\beta}_2$ takes the form
\begin{align*}
&p(Y_{1:T} = y_{1:T}| \bm{X}_{1:T}=\bm{x}_{1:T}, \phi) \propto  \sqrt{\frac{1-\phi^2}{|L_{XX}(\phi)|}} \int  \ d \sigma^2 \bigg( \frac{1}{\sigma^2} \bigg)^{\frac{T-n}{2} + 1} \exp \bigg( - \frac{1}{2 \sigma^2} g(\phi) \bigg),
\end{align*}
where $g(\phi)$ has the form 
\begin{equation*}
g(\phi) = L_{YY}(\phi) - L^\top_{XY} (\phi) \big( L_{XX}(\phi) \big)^{-1} L_{XY}(\phi).
\end{equation*}
The term inside the integral is proportional to a scaled Inv-$\chi^2$ distribution where the degrees of freedom and scale, which we denote by $\nu_p$ and $s_p$ respectively, are given by 
\begin{align*}
\nu_p &= T-n, \quad \quad \quad s_p(\phi) = \sqrt{\frac{g(\phi)}{\nu_p}}.
\end{align*}
This means that performing the integral over $\sigma^2$ and ignoring multiplicative terms that are independent of $\phi$ gives (\ref{supp_likelihood_ar1_prior3_full_likelihood}).

The calculation of the full marginal likelihood (\ref{supp_likelihood_ar1_prior3_full_likelihood}) uses the fact that $\phi \in (-1,1)$. To calculate the marginal likelihood for the unit root process we now calculate the following limit
\begin{equation}
p(Y_{1:T} = y_{1:T}|\bm{X}_{1:T} = \bm{x}_{1:T},  \phi=1) = \lim_{\phi \to 1} p(Y_{1:T} = y_{1:T}|\bm{X}_{1:T} = \bm{x}_{1:T}, \phi). \label{initial_observation_limit}
\end{equation}
To consider the limit we first observe that (\ref{supp_likelihood_ar1_prior3_full_likelihood}) can be written in the form 
\begin{equation}
p(Y_{1:T} = y_{1:T}|\bm{X}_{1:T} = \bm{x}_{1:T}, \phi) \propto \sqrt{\frac{f_1(\phi)}{f_2(\phi)}}, \label{ratio_of_polynomialsInPhi}
\end{equation}
where $f_1 \in \mathcal{P}_m$ and $f_2 \in \mathcal{P}_n$, for some $m,n \in \mathbb{N}$. This can be seen from the fact that the elements of $L_{XX}(\phi)$, $L_{XY}(\phi)$ and $L_{YY}(\phi)$ are quadratic in $\phi$, which means that $|L_{XX}(\phi)|$ is a polynomial in $\phi$ and $L_{XX}^{-1}(\phi)$, assuming it exists, is a matrix whose elements are rational in $\phi$. This means that $g(\phi)$ is a rational function of $\phi$ and it follows that (\ref{supp_likelihood_ar1_prior3_full_likelihood}) can be written in the form (\ref{ratio_of_polynomialsInPhi}). To consider the limit in (\ref{initial_observation_limit}) we now consider the change of variable, $\epsilon = 1 - \phi$, and then consider the limit as $\epsilon \to 0$. For any $\epsilon > 0$ the marginal likelihood can be written in the form   
\begin{equation}
p(Y_{1:T} = y_{1:T}|\bm{X}_{1:T} = \bm{x}_{1:T}, \phi) \propto \sqrt{\frac{h_1(\epsilon)}{h_2(\epsilon)}}, \label{ratio_of_polynomials}
\end{equation}
where $h_1 \in \mathcal{P}_m$ and $h_2 \in \mathcal{P}_n$. We write the coefficients of $h_1$ and $h_2$ as $\{a_i\}_{i=0}^m$ and $\{b_i\}_{i=0}^n$, respectively, where $a_i$ (or $b_i$) is the coefficient of the $i^{\textnormal{th}}$ order term in the corresponding polynomial. The limit as $\epsilon \to 0$ will exist provided that $\min \{i \in \mathbb{N}_n | b_i \ne 0\} \le \min \{i \in \mathbb{N}_m | a_i \ne 0\} $, \emph{i.e.} that the lowest order (non-zero) term in $h_2$ has order less than or equal to the lowest order (non-zero) term in $h_1$. Provided that this property is satisfied then the limit can then be obtained through the (possibly repeated) application L'H\^{o}pital's rule. In practice the polynomials $h_1$ and $h_2$ will be very difficult to calculate and so instead the limit will have to be evaluated numerically, for example through Richardson extrapolation \cite{extrapolation_book}.

\subsection{Higher-Order Autoregressive Residual Processes}\label{supp_section_AR(p)_cointegration_test_p_known}

We now derive the alternative representation (\ref{pthOrderResidualAutoregressiveProcessErrorCorrectionForm}) of the autoregression (\ref{pthOrderResidualAutoregressiveProcess}). Starting with (\ref{pthOrderResidualAutoregressiveProcessErrorCorrectionForm}) we have that 
\begin{equation}
R_t = \rho R_{t-1} + \xi_1 \Delta R_{t-1} + \xi_2 \Delta R_{t-2} + \cdots + \xi_{p-1} \Delta R_{t-(k-1)} + \epsilon_t, \label{supppthOrderResidualAutoregressiveProcessErrorCorrectionForm} 
\end{equation}
where 
\begin{equation*}
\rho = \sum_{i=1}^k \phi_i, \quad \quad \quad \quad \xi_i = - \sum_{j=i+1}^k \phi_j.
\end{equation*}
Using the fact that $\Delta R_{t-1} = R_{t-1} - R_{t-2}$ in (\ref{supppthOrderResidualAutoregressiveProcessErrorCorrectionForm}) gives 
\begin{equation}
R_t = (\rho + \xi_1) R_{t-1} - \xi_1 R_{t-2} + \xi_2 \Delta R_{t-2} + \cdots + \xi_{k-1} \Delta R_{t-(k-1)} + \epsilon_t. \label{supppthOrderResidualAutoregressiveProcessErrorCorrectionFormeq2} 
\end{equation}
It can be seen that $\rho + \xi_1 = \phi_1$, so that (\ref{supppthOrderResidualAutoregressiveProcessErrorCorrectionFormeq2}) takes the form 
\begin{equation}
R_t = \phi_1 R_{t-1} - \xi_1 R_{t-2} + \xi_2 \Delta R_{t-2} + \cdots + \xi_{k-1} \Delta R_{t-(k-1)} + \epsilon_t. \label{supppthOrderResidualAutoregressiveProcessErrorCorrectionFormeq3} 
\end{equation}
In a similar manner, using the fact that $\Delta R_{t-2} = R_{t-2} - R_{t-3}$ in (\ref{supppthOrderResidualAutoregressiveProcessErrorCorrectionFormeq3}), along the fact that $\xi_2-\xi_1 = \phi_2$, gives 
\begin{equation}
R_t = \phi_1 R_{t-1} + \phi_2 R_{t-2} - \xi_2 R_{t-3} + \xi_3 \Delta R_{t-3} + \cdots + \xi_{k-1} \Delta R_{t-(k-1)} + \epsilon_t. \label{supppthOrderResidualAutoregressiveProcessErrorCorrectionFormeq4} 
\end{equation}
This process can now be repeated until one obtains the original form of the autoregression
\begin{equation*}
R_t = \phi_1 R_{t-1} + \phi_2 R_{t-2} + \cdots + \phi_k R_{t-k} + \epsilon_t.
\end{equation*}

We now show that $\rho$ has the form 
\begin{equation}
\rho = (-1)^{k+1} \prod_{i=1}^k (\lambda_i - 1) + 1, \label{supprhoInTermsOfZeros}
\end{equation}
where $\{\lambda_i\}_{i=1}^k$ are the roots of the polynomial $\Pi(z) = z^k - \phi_1 z^{k-1} - \phi_2 z^{k-2} - \cdots - \phi_k$. Firstly, as $\{\lambda_i\}_{i=1}^k$ are the roots of $\Pi$ we have that $\Pi$ can be written in the equivalent form
\begin{align*}
\Pi(z) &= \prod_{i=1}^k (z - \lambda_i) = \sum_{i=0}^k (-1)^{k-i} e_{k-i}(\lambda_1, \lambda_2, ..., \lambda_k) z^i,
\end{align*}
where $\{e_{i}(\lambda_1, \lambda_2, ..., \lambda_k)\}_{i=0}^k$ are the elementary symmetric polynomials given by 
\begin{align*}
e_0 (\lambda_1, \lambda_2, \dots, \lambda_k) = 1, \quad \quad e_1 &(\lambda_1, \lambda_2, \dots, \lambda_k) = \textstyle\sum_{1 \leq i \leq k} \lambda_i, \\
e_2 (\lambda_1, \lambda_2, \dots, \lambda_k) &= \textstyle\sum_{1 \leq i < j \leq k} \lambda_i \lambda_j,
\end{align*} 
and so on. Equating coefficients of $\Pi$ gives 
\begin{equation*}
\phi_i = (-1)^{i+1} e_i(\lambda_1, \lambda_2, \dots, \lambda_k), \quad i \in \mathbb{N}_k.
\end{equation*}
Now $\rho = \sum_{i=1}^k \phi_i$, so that 
\begin{align*}
\rho = \sum_{i=1}^k (-1)^{i+1} e_i(\lambda_1, \lambda_2, \dots, \lambda_k) = \sum_{i=0}^k (-1)^{i+1} e_i(\lambda_1, \lambda_2, \dots, \lambda_k) 1^{k-i} + 1.
\end{align*}
Now 
\begin{equation*}
\sum_{i=0}^k (-1)^{i+1} e_i(\lambda_1, \lambda_2, \dots, \lambda_k) 1^{k-i} = (-1) \prod_{i=1}^k (1 - \lambda_i) = (-1)^{k+1} \prod_{i=1}^k (\lambda_i - 1),
\end{equation*}
so that $\rho$ takes the final form
\begin{equation*}
\rho = (-1)^{k+1} \prod_{i=1}^k (\lambda_i - 1) + 1.
\end{equation*}

We now detail the derivation of the conditional distributions 
\begin{align*}
p(\rho, \bm{\xi}|\bm{Z}_{1:T}=\bm{z}_{1:T}, \sigma^2, \bm{\beta}_2), \quad  \quad  p(\bm{\beta}_2|\bm{Z}_{1:T}=\bm{z}_{1:T}, \rho, \bm{\xi}, \sigma^2), \quad \quad  p(\sigma^2|\bm{Z}_{1:T}=\bm{z}_{1:T}, \rho, \bm{\xi}, \bm{\beta}_2).
\end{align*}
Firstly, the conditional distribution, $p(\rho, \bm{\xi}|\bm{Z}_{1:T}=\bm{z}_{1:T}, \sigma^2, \bm{\beta}_2)$ is given by
\begin{align*}
p(\rho, \bm{\xi}|\bm{Z}_{1:T}=\bm{z}_{1:T}, \sigma^2, \bm{\beta}_2) \propto p(Y_{k+1:T}=y_{k+1:T}| \bm{X}_{1:T} = \bm{x}_{1:T}, \rho, \bm{\xi}, \sigma^2, \bm{\beta}).
\end{align*}
Given (\ref{pthOrderResidualAutoregressiveProcessErrorCorrectionForm}) this conditional distribution takes the form
\begin{equation}
p(\rho, \bm{\xi}|\bm{Z}_{1:T} = \bm{z}_{1:T}, \sigma^2, \bm{\beta}) \propto \prod_{t=k+1}^T \mathcal{N} \bigg( R_t \bigg| \rho R_{t-1} + \sum_{i=1}^{k-1} \xi_i \Delta R_{t-i}, \sigma^2 \bigg). \label{ARp_conditional_dist1}
\end{equation}
Expanding the exponent of (\ref{ARp_conditional_dist1}) and completing the square in $\begin{bmatrix} \rho & \bm{\xi} \end{bmatrix}^\top$ gives 
\begin{equation}
\rho, \bm{\xi} | \dots \sim \mathcal{N} \bigg( \big( X_{\rho, \bm{\xi}} X_{\rho, \bm{\xi}}^\top \big)^{-1} X_{\rho, \bm{\xi}} Y_{\rho, \bm{\xi}}, \sigma^2  \big( X_{\rho, \bm{\xi}} X_{\rho, \bm{\xi}}^\top \big)^{-1} \bigg).
\end{equation}
The terms $X_{\rho, \bm{\xi}}$ and $Y_{\rho, \bm{\xi}}$ are a $k \times (T-k)$ matrix and a $(T-k)$-dimensional vector, respectively, and are given by 
\begin{equation*}
X_{\rho, \bm{\xi}} = \begin{bmatrix}[cccc] R_k & R_{k+1} & \dots & R_{T-1} \\ \Delta R_k & \Delta R_{k+1} & \dots & \Delta R_{T-1} \\ \vdots & \vdots & \ddots & \vdots \\ \Delta R_2 & \Delta R_3 & \dots & \Delta R_{T+1-k} \end{bmatrix}, \qquad \qquad Y_{\rho, \bm{\xi}} = \begin{bmatrix} R_{k+1} \\ R_{k+2} \\ \vdots \\ R_T \end{bmatrix}.
\end{equation*}

The conditional distribution, $p(\bm{\beta}_2|\bm{Z}_{1:T}=\bm{z}_{1:T}, \rho, \bm{\xi}, \sigma^2)$, is given by 
\begin{align*}
p(\bm{\beta}_2|\bm{Z}_{1:T}=\bm{z}_{1:T}, \rho, \bm{\xi}, \sigma^2) &\propto p(Y_{k+1:T}=y_{k+1:T}| \bm{X}_{1:T} = \bm{x}_{1:T}, \rho, \bm{\xi}, \sigma^2, \bm{\beta}_2),
\end{align*}
Given (\ref{pthOrderResidualAutoregressiveProcessErrorCorrectionForm}) this conditional distribution takes the form
\begin{equation}
p(\bm{\beta}_2|\bm{Z}_{1:T}=\bm{z}_{1:T}, \rho, \bm{\xi}, \sigma^2) \propto \prod_{t=k+1}^T \mathcal{N} \bigg( R_t \bigg| \rho R_{t-1} + \sum_{i=1}^{k-1} \xi_i \Delta R_{t-i}, \sigma^2 \bigg). \label{ARp_conditional_dist3}
\end{equation}
To obtain a closed form for $p(\bm{\beta}_2|\bm{Z}_{1:T}=\bm{z}_{1:T}, \rho, \bm{\xi}, \sigma^2)$ we need to write the exponent of (\ref{ARp_conditional_dist3}) as a quadratic in the regression parameters, where we then complete the square in the regression parameters to obtain a Gaussian sampling distribution. To do so we first note that, for each $t \in \mathbb{N}_T$, we have
\begin{equation*}
R_t = Y_t - \bm{\beta}^\top_2 \bm{X}_t, \quad \quad \quad \Delta R_t = \Delta Y_t - \bm{\beta}^\top_2 \Delta \bm{X}_t.
\end{equation*}
This means that (\ref{ARp_conditional_dist3}) can be written in the equivalent form
\begin{align*}
& p(\bm{\beta}_2|\bm{Z}_{1:T}=\bm{z}_{1:T}, \rho, \bm{\xi}, \sigma^2) \\ 
& \propto \prod_{t=p+1}^T \mathcal{N} \bigg( Y_t - \bigg( \rho Y_{t-1} + \sum_{i=1}^{p-1} \xi_i \Delta Y_{t-i} \bigg) \bigg| \bm{\beta}^\top_2 \bigg( \bm{X}_t - \bigg( \rho \bm{X}_{t-1} + \sum_{i=1}^{p-1} \xi_i \Delta \bm{X}_{t-i} \bigg) \bigg), \sigma^2 \bigg). \label{ARp_conditional_dist3_form2}
\end{align*}
Expanding the exponent and completing the square in $\bm{\beta}_2$ gives 
\begin{equation}
\bm{\beta}_2 | \dots \sim \mathcal{N} \bigg( \big( X_{\bm{\beta}_2} X_{\bm{\beta}_2}^\top \big)^{-1} X_{\bm{\beta}_2} Y_{\bm{\beta}_2}, \sigma^2 \big( X_{\bm{\beta}_2} X_{\bm{\beta}_2}^\top \big)^{-1} \bigg).
\end{equation}
The terms $X_{\bm{\beta}_2}$ and $Y_{\bm{\beta}_2}$ are a $n-1 \times (T-k)$ matrix and a $(T-k)$-dimensional vector, respectively, where the $t^{\textnormal{th}}$ column of $X_{\bm{\beta}_2}$ is given by $\rho \bm{X}_{t+k-1} + \sum_{i=1}^{k-1} \xi_i \Delta \bm{X}_{t+k-i}$. The elements of $Y_{\bm{\beta}_2}$ are defined in a similar manner. 

The final conditional distribution, $p(\sigma^2|\bm{Z}_{1:T} = \bm{z}_{1:T}, \rho, \bm{\xi}, \bm{\beta}_2)$, takes the form 
\begin{align*}
p(\sigma^2| \bm{Z}_{1:T} = \bm{z}_{1:T}, \rho, \bm{\xi}, \bm{\beta}) \propto p(Y_{k+1:T}=y_{k+1:T}| \bm{X}_{1:T} = \bm{x}_{1:T}, \rho, \bm{\xi}, \sigma^2, \bm{\beta}) p(\sigma^2).
\end{align*}
Given the prior, $p(\sigma^2) = \frac{1}{\sigma^2}$, this conditional distribution takes the form 
\begin{equation*}
p(\sigma^2| \bm{Z}_{1:T} = \bm{z}_{1:T}, \rho, \bm{\xi}, \bm{\beta}) \propto \bigg( \frac{1}{\sigma^2} \bigg)^{\frac{T-k}{2} + 1}  \exp \bigg( -\frac{1}{2 \sigma^2} \sum_{t=k+1}^T \big( R_t - \big( \rho R_{t-1} + \sum_{i=1}^{k-1} \xi_i \Delta R_{t-i}  \big) \big)^2 \bigg),
\end{equation*}
which is an scaled-inverse-$\chi^2$ distribution with degrees of freedom, $\nu = T-k$, and scale parameter,
\begin{equation*}
\tau^2 = \frac{1}{T-k} \bigg( \sum_{t=k+1}^T \big( R_t - \big( \rho R_{t-1} + \sum_{i=1}^{k-1} \xi_i \Delta R_{t-i}  \big) \big)^2 \bigg).
\end{equation*}
The use of the notation $\{R_t\}_{t=1}^T$ in $\tau^2$, $X_{\rho, \bm{\xi}}$ and $Y_{\rho, \bm{\xi}}$ refers to the residual process induced by the current sample of the regression coefficients.

\subsection{Autoregressive Residual Processes with Unknown Order}\label{supp_section_AR(p)_cointegration_test_p_unknown}

In this section we derive the acceptance ratio $A((k, \phi_{1:k}) \to (k', \phi'_{1:k'}))$ for the proposal distributions considered in \secref{section_AR(p)_cointegration_test_p_unknown}. We start by noting that $A((k, \phi_{1:k}) \to (k', \phi'_{1:k'}))$ has the general form
\begin{equation}
A((k, \phi_{1:k}) \to (k', \phi'_{1:k'})) = \frac{p(k', \phi'_{1:k'}|\bm{\pi})}{p(k, \phi_{1:k}|\bm{\pi})} \frac{q(k, \phi_{1:k}|k',\phi'_{1:k'}, \bm{\pi})}{q(k', \phi'_{1:k'}|k,\phi_{1:k},\bm{\pi})}, \label{RJMCMC_acceptance_probability1}
\end{equation}
where $q$ is the proposal distribution for between-model moves, $p(k', \phi'_{1:k'}|\bm{\pi})$ is the posterior density and $\bm{\pi}$ denotes all remaining variables in the conditioning set, which in this case corresponds to the observed collection of time-series, the regression coefficients and the variance of the white noise process. The standard Jacobian term is absent from (\ref{RJMCMC_acceptance_probability1}) are we are proposing moves directly in the new parameter space.

We now consider the specific form for (\ref{RJMCMC_acceptance_probability1}) given the proposal distributions considered in \secref{section_AR(p)_cointegration_test_p_unknown}. In particular, we shall show that (\ref{RJMCMC_acceptance_probability1}) takes the form 
\begin{equation}
A((k, \phi_{1:k}) \to (k', \phi'_{1:k'})) = \frac{q(k|k')}{q(k'|k)} \sqrt{ \frac{ \big| 2 \pi \sigma^2 \big( X_{\rho, \bm{\xi}}(k') X_{\rho, \bm{\xi}}^\top(k') \big)^{-1} \big|}{ \big| 2 \pi \sigma^2 \big( X_{\rho, \bm{\xi}}(k) X_{\rho, \bm{\xi}}^\top(k) \big)^{-1} \big|}} \exp \bigg(\frac{1}{2} \big( C (k') - C(k) \big) \bigg), \label{RJMCMC_acceptance_probability2}
\end{equation}
where 
\begin{equation*}
C(k) = Y^\top_{\rho, \bm{\xi}}(k) X^\top_{\rho, \bm{\xi}}(k)  \big( X_{\rho, \bm{\xi}}(k) X_{\rho, \bm{\xi}}^\top(k) \big)^{-1} X_{\rho, \bm{\xi}}(k) Y_{\rho, \bm{\xi}}(k).
\end{equation*}
The notation $X_{\rho, \bm{\xi}}(k)$ is used to denote the matrix from \secref{section_AR(p)_cointegration_test_p_known} for the case where the model is of size $k$ and the first $\max (k,k')$ observations are conditioned. Similar notation is used for the other terms in (\ref{RJMCMC_acceptance_probability2}). Note that (\ref{RJMCMC_acceptance_probability2}) is independent of the autoregression parameters, so that it is only necessary to sample the autoregression parameters after a move has been accepted.


We now derive the form of the acceptance ratio (\ref{RJMCMC_acceptance_probability2}), where this derivation closely follows a similar derivation in \cite{Troughton97}. We shall consider the acceptance ratio of the proposed jump $(k, \phi_{1:k}) \to (k', \phi'_{1:k'})$. Firstly, the candidate's identity \cite{besag-1989} gives 
\begin{equation}
p(k|\bm{Z}_{1:T} = \bm{z}_{1:T}, \bm{\beta}_2, \sigma^2) = \frac{p(k, \phi_{1:k}|\bm{Z}_{1:T} = \bm{z}_{1:T}, \bm{\beta}_2, \sigma^2)}{p(\phi_{1:k}|\bm{Z}_{1:T} = \bm{z}_{1:T}, k, \bm{\beta}_2, \sigma^2)}. \label{candidateIndentity}
\end{equation}
As the proposal distribution for the autoregressive parameters is given by 
\begin{equation}
q(\phi_{1:k}|k, k',\phi'_{1:k'}, \bm{Z}_{1:T} = \bm{z}_{1:T}, \bm{\beta}_2, \sigma^2) = p(\phi_{1:k}|\bm{Z}_{1:T} = \bm{z}_{1:T}, k, \bm{\beta}_2, \sigma^2), \label{suppAccept}
\end{equation}
then an application of (\ref{candidateIndentity}) into (\ref{RJMCMC_acceptance_probability1}) gives
\begin{equation}
A = \frac{q(k|k')}{q(k'|k)} \frac{p(k'|\bm{Z}_{1:T} = \bm{z}_{1:T}, \bm{\beta}_2, \sigma^2)}{p(k|\bm{Z}_{1:T} = \bm{z}_{1:T}, \bm{\beta}_2, \sigma^2)}, \label{suppRJMCMC_acceptance_probability1}
\end{equation}
where $q(k'|k)$ denotes the proposal distribution used to propose moves in the order of the residual process. So to calculate the acceptance ratio it is sufficient to calculate the posterior of the model order and the proposal distribution for jumps in the model order. The later calculation is given by the discretised Laplacian, so we detail the calculation of the posterior of the model order. Firstly, we have that 
\begin{equation*}
p(k|\bm{Z}_{1:T} = \bm{z}_{1:T}, \bm{\beta}_2, \sigma^2) \propto p(k) p(Y_{k+1:T} = y_{k+1:T}|\bm{X}_{1:T} = \bm{x}_{1:T}, k, \bm{\beta}_2, \sigma^2),
\end{equation*}
where we have considered the conditional likelihood. As we are considering a flat prior on the parameters of the autoregression we have that 
\begin{equation*}
p(k|\bm{Z}_{1:T} = \bm{z}_{1:T}, \bm{\beta}_2, \sigma^2) \propto p(k) \int d \phi_{1:k} \ p(Y_{k+1:T} = y_{k+1:T}|\bm{X}_{1:T} = \bm{x}_{1:T}, k, \phi_{1:k}, \bm{\beta}_2, \sigma^2).
\end{equation*}
After some standard manipulations the conditional likelihood can be written in the form
\begin{equation*}
p(Y_{k+1:T} = y_{k+1:T}|\bm{X}_{1:T} = \bm{x}_{1:T}, k, \phi_{1:k}, \bm{\beta}_2, \sigma^2) = \mathcal{N} \big( Y_{\rho, \bm{\xi}} \big| X_{\rho, \bm{\xi}}^\top \phi_{1:k}, \sigma^2 I_{T-k} \big),
\end{equation*}
which after completing the square \emph{w.r.t.} the parameters of the autoregression gives
\begin{align*}
p(Y_{k+1:T}  = y_{k+1:T} &| \bm{X}_{1:T} = \bm{x}_{1:T}, k, \phi_{1:k}, \bm{\beta}_2, \sigma^2) \\
&= (2 \pi \sigma^2)^{-\frac{T-k}{2}} \exp \bigg( -\frac{1}{2 \sigma^2} \big( Y^\top_{\rho, \bm{\xi}} Y_{\rho, \bm{\xi}} - Y^\top_{\rho, \bm{\xi}} X^\top_{\rho, \bm{\xi}} \big( X_{\rho, \bm{\xi}} X^\top_{\rho, \bm{\xi}} \big)^{-1} X_{\rho, \bm{\xi}}  Y_{\rho, \bm{\xi}} \big) \bigg) \\
& \times \sqrt{ \bigg| 2 \pi \sigma^2 \big( X_{\rho, \bm{\xi}} X^\top_{\rho, \bm{\xi}} \big)^{-1} \bigg|} \mathcal{N} \bigg(\phi_{1:k} \bigg| \big( X_{\rho, \bm{\xi}} X^\top_{\rho, \bm{\xi}} \big)^{-1} X_{\rho, \bm{\xi}}  Y_{\rho, \bm{\xi}}, \sigma^2 \big( X_{\rho, \bm{\xi}} X^\top_{\rho, \bm{\xi}} \big)^{-1} \bigg).
\end{align*}
The integral over the parameters of the autoregression can now be performed to give
\begin{align*}
p(k|\bm{Z}_{1:T} = \bm{z}_{1:T}, \bm{\beta}_2, \sigma^2) &\propto p(k) (2 \pi \sigma^2)^{-\frac{T-k}{2}} \sqrt{ \bigg| 2 \pi \sigma^2 \big( X_{\rho, \bm{\xi}} X^\top_{\rho, \bm{\xi}} \big)^{-1} \bigg|} \\
& \times \exp \bigg( -\frac{1}{2 \sigma^2} \big( Y^\top_{\rho, \bm{\xi}} Y_{\rho, \bm{\xi}} - Y^\top_{\rho, \bm{\xi}} X^\top_{\rho, \bm{\xi}} \big( X_{\rho, \bm{\xi}} X^\top_{\rho, \bm{\xi}} \big)^{-1} X_{\rho, \bm{\xi}}  Y_{\rho, \bm{\xi}} \big) \bigg).
\end{align*}
By considering the conditional likelihood, where we condition on the first $\max (k, k')$ observations, for both terms in (\ref{suppAccept}) we obtain the acceptance ration (\ref{RJMCMC_acceptance_probability2}), where we have used the fact that $p(k) = k^{-1}_{\textnormal{max}}$.

\subsection{Experiments}\label{supp_material_experiments}

We now detail the procedure used to generate the synthetic data in the cointegration classification experiment. In this experiment $R_t$ was modelled with an autoregressive process of known order, where in the first experiment the order was one, while in the second experiment the order was three. The residual process was either stationary or had a unit root present, where these two possibilities were considered with equal probability. When $R_t$ was stationary the autoregressive parameters were uniformly sampled from a subspace of the space of stationary models, where we only considered stationary models with positive parameters and with at least one root of $\Pi$ greater than $0.8$ in magnitude. This was done to ensure that the stationary process was sufficiently close to a unit root process so that the resulting classification test was not trivial. When there was a unit root present in $R_t$ we first generated the parameters of $\Delta R_t$, from the same subspace of stationary models as before, and then obtained the autoregressive parameters for $R_t$ from $\Delta R_t$. Rejection sampling was used to perform the sampling. The variance of the white noise process was set to one. We considered the relation between two univariate time-series, $X_t$ and $Y_t$, where $Y_t$ was regressed upon $X_t$. We included an intercept in the regression and uniformly sampled the regression parameters from the interval $[0.0, 5.0]$. The regressor time-series, $X_{1:T}$, was generated through a random walk, while the regressand time-series was generated through the relation (\ref{cointegrationRelationshipEquation}).

We now detail the procedure used to generate the synthetic data in the RJ-MCMC model determination experiment. In this experiment $R_t$ was modelled with an autoregressive process, where the order of the model was uniformly selected from models up to and including third-order models. The residual process was either stationary or had a unit root present, where these two possibilities were considered with equal probability. When $R_t$ was stationary the autoregressive parameters were uniformly sampled from a subspace of the space of stationary models, where we only considered stationary models with positive parameters and with at least one root of $\Pi$ greater than $0.8$ in magnitude. When there was a unit root present in $R_t$ we first generated the parameters of $\Delta R_t$, from the same subspace of stationary models as before, and then obtained the autoregressive parameters for $R_t$ from $\Delta R_t$. Rejection sampling was used to perform the sampling. The variance of the white noise process was set to one. We considered the relation between two univariate time-series, $X_t$ and $Y_t$, where $Y_t$ was regressed upon $X_t$. We included an intercept in the regression and uniformly sampled the regression parameters from the interval $[0.0, 5.0]$. The regressor time-series, $X_{1:T}$, was generated through a random walk, while the regressand time-series was generated through the relation (\ref{cointegrationRelationshipEquation}).

\end{document}